\documentclass[11pt]{amsart}
\usepackage{amsbsy,amssymb,amsmath,amsthm,amscd,amsfonts,latexsym,amstext,delarray,
amsmath,graphicx} 
\usepackage[margin=1in]{geometry}
\usepackage{color}
\input xypic

\newtheorem{thm}{Theorem}[section]
\newtheorem{prop}[thm]{Proposition}
\newtheorem{cor}[thm]{Corollary}
\newtheorem{lem}[thm]{Lemma}

\newtheorem{rem}[thm]{Remark}
\newtheorem{ex}[thm]{Example}

\numberwithin{equation}{section}

\def\F{{\mathbb F}}
\def\Q{{\mathbb Q}}
\def\Z{{\mathbb Z}}

\def\R{{\mathbb R}}
\def\C{{\mathbb C}}
\renewcommand\P{{\mathbb P}}
\def\H{{\mathbb H}}

\def\P{{\mathbb P}}

\def\fA{{\mathfrak A}}

\def\cA{{\mathcal A}}

\def\cL{{\mathcal L}}

\def\cN{{\mathcal N}}
\def\cO{{\mathcal O}}

\def\cR{{\mathcal R}}
\def\cS{{\mathcal S}}
\def\cT{{\mathcal T}}

\def\cW{{\mathcal W}}

\def\bK{{\mathbb K}}

\def\fm{{\mathfrak m}}

\title{Multifractals, Mumford curves, and Eternal Inflation}
\author{Matilde Marcolli and Nicolas Tedeschi}
\address{Mathematics Department, Caltech, 1200 E. California Blvd. Pasadena, CA 91125, USA}
\email{matilde@caltech.edu}
\email{nicot@caltech.edu}
\date{}

\begin{document}
\maketitle

\begin{abstract}
We relate the Eternal Symmetree model of Harlow, Shenker, Stanford, and Susskind
to constructions of stochastic processes related to quantum statistical mechanical
systems on Cuntz--Krieger algebras. We extend the eternal inflation model from the
Bruhat--Tits tree to quotients by $p$-adic Schottky groups, again using quantum
statistical mechanics on graph algebras.
\end{abstract}

\tableofcontents

\section{Introduction}

A model of eternal inflation based on a tree structure was developed
recently by Harlow, Shenker, Stanford, and Susskind, see \cite{HSSS}
and \cite{Suss}, based on $p$-adic Bruhat--Tits trees.

\smallskip

We revisit the model here from the point of view of fractal geometry
and noncommutative geometry, using constructions of stochastic
processes, multifractal measures and wavelets, associated to Cuntz
and Cuntz--Krieger algebras, \cite{ManMar}, \cite{MaPa}, as well 
as a operator algebraic methods applied to the geometry of 
$p$-adic Mumford curves, as previously developed in \cite{CMR}.

\smallskip

In particular, we show that the type of stochastic process considered
in the eternal inflation model of \cite{HSSS}, in the case of a particular
class of ``pruning methods" associated to subshifts of finite type,
can be obtained from the KMS equilibrium states of a quantum statistical 
mechanical system on a noncommutative operator algebra associated 
to the pruned tree.  In particular this implies that what plays the role of
the proper time in the model, which gives the discrete evolution of the
stochastic process, in turn can be seen as depending on an internal
notion of time evolution acting on the creation and annihilation
operators given by the generators of the noncommutative algebra
associated to the graph. The propagators in the correlation functions
for the multiverse fields of \cite{HSSS} in turn provide a measure of
autocorrelation for wavelets on fractals arising from the construction
of the multifractal measure (as in \cite{DuJo}, \cite{MaPa}) on the
boundary of the tree, which determined the stochastic process.

\smallskip

We also show how one can extend the eternal inflation model
from the case of the $p$-adic Bruhat--Tits tree to infinite graphs
given by quotients of the Bruhat--Tits tree by a $p$-adic Schottky
group. These graphs have boundary at infinity given by a $p$-adically
uniformized Mumford curve, and they consist of a central finite graph
(the dual graph of the closed fiber of the minimal smooth model
of the curve) with infinite trees sticking out of its vertices.
We show that one can consistently interpolate random processes on the external
trees of the type considered in \cite{HSSS} to a stochastic process on the
entire graph, which is constructed using KMS weights of an associated
graph $C^*$-algebra. The picture that emerges is one where the dynamics
of the eternal inflation model can remain trapped inside a bounded region
given by the finite graph, or wander off into one of the trees, where it
reproduces the model of \cite{HSSS}.

\smallskip
\section{The Eternal Symmetree model}

The {\em Eternal Symmetree} of \cite{HSSS} is a discretized model 
of eternal inflation. In this model, a multiverse landscape arises
through a stochastic process describing the likelihood of transitions
between different types of vacua, labeled by the letters of a finite
alphabet $\fA=\{ 0 ,\ldots, p-1 \}$. Each vacuum represents a collection
of microstates with assigned entropies $S=(S_a)_{a\in \fA}$.

\smallskip

The causal future of a node in the tree is the oriented subtree that
branches off from that node to the boundary at infinity. Time is discretized,
with the proper time between two adjacent nodes being given by
a fixed amount, the inverse of the Hubble constant, which can vary
with the label attached to the edge. 

\smallskip

A collection of {\em multiverse fields} is described in \cite{HSSS}, with correlation
functions expressed in terms of the data of the stochastic process
and of the $p$-adic distance on the Bruhat--Tits tree. The group 
${\rm PGL}_2(\Q_p)$ of isometries of the Bruhat--Tits tree of $\Q_p$ 
acts as conformal symmetries. The symmetry is broken if the
tree is suitably ``pruned", giving rise to {\em terminal vacua}.
This alters the form of the the correlation functions, leading to the
emergence of a ``fractal-flow" arrow-of-time, see \cite{Suss}.

\smallskip

We first describe how to reinterpret the case without terminal vacua
in terms of multifractal measures and operator algebra arising from
representations of the Cuntz algebras of \cite{Cu}. We show, in
particular, that the construction of the {\em multiverse fields} in the
Eternal Symmetree model is closely related to the construction of
\cite{DuJo} (see also \cite{ManMar}) of stochastic processes and wavelets 
on the Cantor sets dual to the maximal abelian subalgebra of the
Cuntz algebra. 

\smallskip

We will then consider the case with terminal vacua, where we
focus on pruning of the tree obtained through an admissibility
condition on adjacent edges. We will show that the model can
be reinterpreted as passing from Cuntz algebras to Cuntz--Krieger
algebra, where once again one can relate the multiverse fields to
stochastic processes, multifractal measures and wavelets on the 
associated Cantor sets, as in \cite{MaPa}.

\begin{center}
\begin{figure}
\includegraphics[scale=.8]{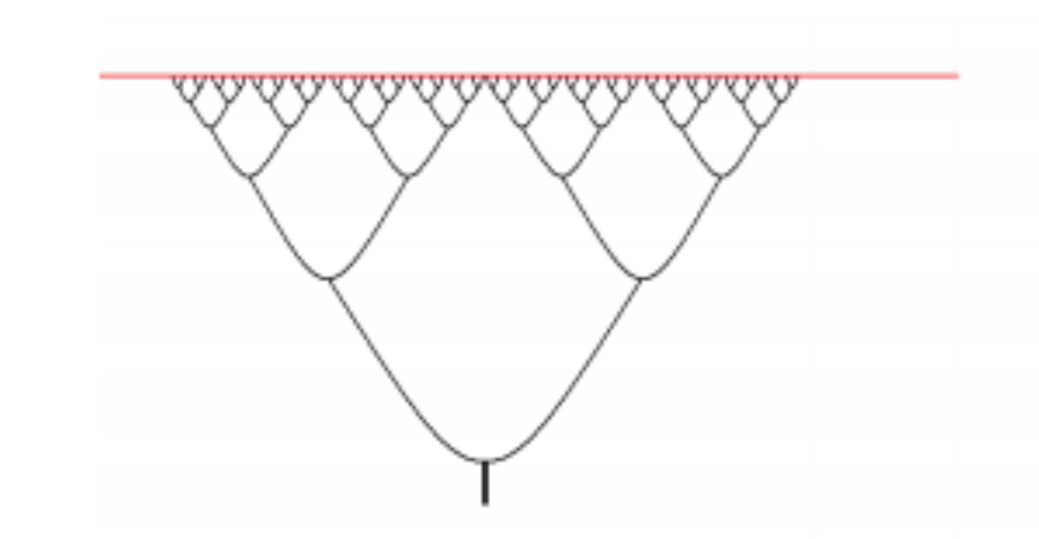}
\includegraphics[scale=.8]{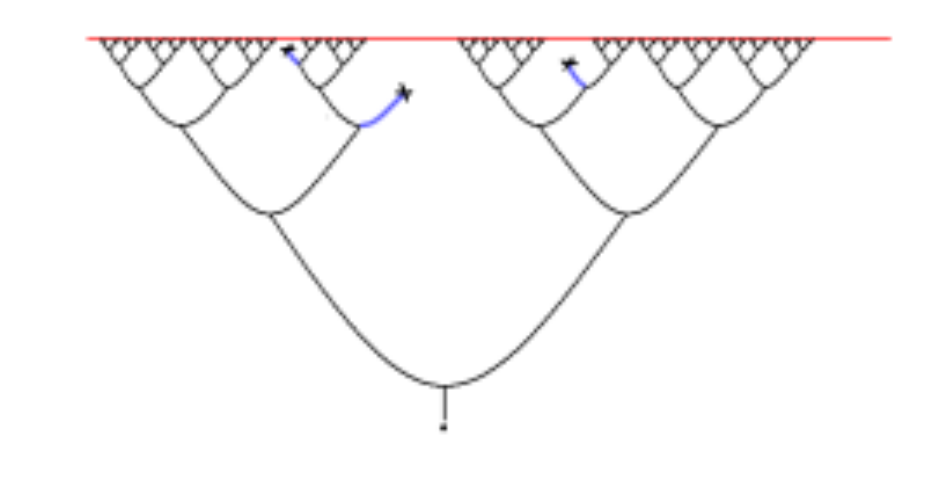}
\caption{The Eternal Symmetree without and with pruning, as
illustrated in \cite{HSSS}.}
\end{figure}
\end{center}

\smallskip
\subsection{Bruhat--Tits trees}

Let $\cT$ be a uniform infinite tree with vertices of valence $q+1$, with $q\geq 2$.  We are
interested in particular in the case where $\cT$ is the Bruhat--Tits tree 
of ${\rm PGL}_2(\bK)$, with $\bK$ a finite extension of $\Q_p$.  In this case
the integer $q=p^r$ is the cardinality of the residue field of $\bK$. A ray is a half-infinite
path without backtracking and an infinite geodesic is an infinite path 
without backtracking. We denote by $\partial\cT$ the boundary at infinity of the tree, 
which is the set of equivalence classes of rays, where two rays are equivalent if they
have an infinite number of vertices in common. Any choice of two
distinct points on the boundary determines a unique infinite geodesic in $\cT$
that connects them. In the case of the Bruhat--Tits tree, the boundary is
identified with $\P^1(\bK)$. We refer the reader to \cite{GePu}, \cite{Man}, \cite{Mum} 
for a detailed exposition of the $p$-adic geometry of Bruhat--Tits trees and their quotients. 

\smallskip

The choice of a coordinate function $z$ on $\P^1(\bK)=\partial \cT$ corresponds to
fixing the choice of points $\{ 0, 1, \infty \}$ in $\P^1(\bK)$. This in turn determines
a unique choice of a vertex $v_0$ of the tree $\cT$, as the unique origin of three
non-overlapping rays with endpoints $\{ 0, 1, \infty \}$. 
Let $v_0$ be a the base vertex in the tree $\cT$ obtained in this way. 
We choose an orientation
of the tree $\cT$ with all the edges pointing outwards from $v_0$, so that at each vertex
$v\neq v_0$ we have one incoming and $q$ outgoing edges. 

\smallskip

In particular, having fixed a projective coordinate on $\P^1(\bK)$ and a corresponding
base vertex in $\cT$ we have a subtree $\cT'$ of the Bruhat--Tits tree $\cT$, with root $v_0$,
whose boundary $\partial \cT' =\cO_{\bK}$ consists of the integers of $\bK$, the 
$p$-adic integers $\Z_p$ in the case where $\bK=\Q_p$. This is the tree considered
in the Eternal Symmetree model of  \cite{HSSS}, \cite{Suss}, where the subtree $\cT'$
of the Bruhat--Tits tree is referred to as the Bethe tree.

\smallskip

This admits an equivalent description in terms of $\omega$-languages, 
which will be useful in the following.

\smallskip
\subsection{$\omega$-languages}

Suppose given a finite alphabet $\fA$ with $\# \fA=q$ with $q\geq 2$. 
Let $\cW_{\fA}$ denote the union $\cW_{\fA}^\star=\cup_{k=0}^\infty \cW_{\fA,k}$
of the sets $\cW_{\fA,k}$ of length $k$ in the alphabet $\fA$. For $k=0$,
$\cW_{\fA,0}$ consists of the empty word $\epsilon=\emptyset$. We denote
by $\cW_\fA^\omega$ the set of all infinite words $a=a_0 a_1 \cdots a_n \cdots$
with $a_k \in \fA$. A {\em language} $\cL$ is a subset of $\cW_{\fA}^\star$ and
an $\omega$-language $\cL^\omega$ is a subset of $\cW_{\fA}^\omega$.

\smallskip

The shift operator $\sigma: \cW_{\fA}^\omega \to \cW_{\fA}^\omega$ is defined
as the map $\sigma: a_0 a_1 \cdots a_n \cdots \mapsto a_1 a_2 \cdots a_{n+1} \cdots $
that shifts the sequence one step to the left and drops the first letter. We require that
the $\omega$-languages $\cL^\omega$ we consider are shift-invariant, in the sense
that if an infinite word $a\in \cW_{\fA}^\omega$ is in $\cL^\omega$, then its shifted
image $\sigma(a)$ is also in $\cL^\omega$.

\smallskip
\subsection{Subshifts of finite type}

We consider in particular $\omega$-languages that are obtained by
imposing an admissibility condition on successive letters in infinite
words in $\cW_{\fA}^\omega$. These have the properties of being
shift-invariant. In terms of the dynamical system defined by the shift
map, they correspond to {\em subshifts of finite type}. 

\smallskip

These are determined by assigning an {\em admissibility matrix} $A=(A_{ab})_{a,b\in \fA}$ 
with entries in $\{ 0,1 \}$. The corresponding $\omega$-languages $\cL_A^\omega$
consists of {\em admissible} infinite words in $\cW_{\fA}^\omega$, namely
those infinite where subsequent letters satisfy the condition that the corresponding
entry of the matrix $A$ is non-zero, 
$$ \cL_A^\omega= \{ a_0 a_1 \cdots a_n \cdots \,|\,  A_{a_k a_{k+1}}=1, \, \forall k\geq 0 \}. $$

\smallskip

Both the space of infinite words $\cW_{\fA}^\omega$ and the subspace $\cL_A^\omega$
can be topologized as  Cantor sets, with a basis of clopen sets given by the cylinders
$\Lambda(w)$, where $\Lambda$ is either $\cW_{\fA}^\omega$
or $\cL_A^\omega$. These are the sets of all infinite words in $\Lambda$ that start
with an assigned (admissible) word $w$ of finite length. 
The shift map $\sigma$ is a continuous dynamical system with respect to this topology.

\smallskip
\subsection{Terminal vacua and subshifts of finite type}

Let $\cT' \subset \cT$ be the Bethe tree, that is, the rooted tree 
with $\partial \cT'=\cO_{\bK}$, as above.

\smallskip

\begin{lem}\label{treeCantor}
The boundary at infinity $\partial \cT'$ of the Bethe tree $\cT'$ can
be equivalently described as the Cantor set of infinite words $\cW_{\fA}^\omega$
in an alphabet $\fA=\F_q$, identified (as a set) with the residue field of $\bK$. 
\end{lem}

\smallskip

\proof
For simplicity, we look at the case $\bK=\Q_p$. The case of finite extensions
is analogous. The $p$-adic integers in $\Z_p$ can be written as infinite series
$x=\sum_{k=0}^\infty x_k p^k$, in powers of $p$, with coefficients in $\{ 0, \ldots, p-1 \}$.
This corresponds to labeling the outgoing edges at each vertex of $\cT'$ 
with a set of labels $\{ e_i \}_{i=0,\ldots, p-1}$. Thus, one can identify rays starting 
at $v_0$ with arbitrary infinite words in the alphabet $\fA=\{ e_i \}_{i=0,\ldots, p-1}$. 
\endproof

\smallskip

The action of the shift operator is related to the notion of proper time 
in the Eternal Symmetree model, which corresponds to the discretized
movement towards one of the next adjacent nodes in the forward
direction along the tree.

\smallskip

We see then that introducing an admissibility condition $A=(A_{ab})$ as above
corresponds to a way of pruning the tree $\cT'$. Namely, any ray of $\cT'$ that
contains a non-admissible consecutive pair of edges $e_k e_{k+1}$ is removed
from the tree, by cutting the branch at the place where the first non-admissible 
pair occurs, coming out of the root. This provides a mechanism that creates 
terminal vacua.

\smallskip

In the Eternal Symmetree model, more general mechanisms for
pruning the tree  $\cT'$ are considered, which do not necessarily correspond to
admissibility conditions defined by a matrix $A$. These other pruning
methods will give rise to more general kinds of $\omega$-languages,
which are not necessarily shift invariant. We focus here only on pruning
defined by admissibility conditions determined by a matrix $A$, as
these will be directly related to an important class of operator algebras,
as we show in the following section.

\section{Multifractal measures via quantum statistical mechanics}

In this section we reinterpret the stochastic process of the
Eternal Symmetree model in terms of multifractal measures
related to representations of Cuntz--Krieger algebras.

\smallskip
\subsection{Entropies and stochastic processes on the Eternal Symmetree}

In the eternal inflation model of \cite{HSSS}, the letters of the alphabet $\fA$
correspond to ``color" labels for the different types of vacua, with each color
corresponding to a collection of microstates. These have associated entropies 
given by a collection $\{ S_a \}_{a\in \fA}$. 
At each node there are probabilities $\gamma_{ab}$ 
of transition from an incoming color $a$ to an outgoing color $b$.
These measure the probability of tunneling between vacua of 
different types.
A {\em detailed balance} condition of microscopic reversibility is
imposed on the probabilities $\gamma_{ab}$, of the form
\begin{equation}\label{detbal}
 \frac{\gamma_{ab}}{\gamma_{ba}}=e^{S_a-S_b} .
\end{equation}
The detailed balance condition is expressed in \cite{HSSS} through
a real symmetric matrix $M$ such that $\gamma_{ab}=M_{ab} e^{S_a}$.

\smallskip

In the case without terminal vacua, a stochastic process is 
constructed out of these data, with
$P_a(k)$ the probability of obtaining a vacuum of type $a\in \fA$ 
after $k$ steps from the root vertex in $\cT'$. These probabilities
are written as $P_a = e^{S_a/2} \Phi_a$, with the $\Phi_a$
satisfying the process
\begin{equation}\label{Phiproc}
\Phi(k+1)=\cS \, \Phi(k),
\end{equation}
where $\cS$ is a positive stochastic matrix with Perron--Frobenius
eigenvalue $\lambda_{\cS}=1$ and positive Perron--Frobenius
eigenvector $v_{\cS}$. 

\smallskip

In \cite{HSSS} the matrix $\gamma_{ab}$ is in turn related to the matrix $\cS_{ab}$
by $\cS=Z^{-1} G Z$ with $Z$ the diagonal matrix with entries $e^{S_a/2}$
and with $G_{ab}=\delta_{ab}-\sum_c \gamma_{ca}\delta_{ab} + \gamma_{ab}$.

\smallskip

We will now reinterpret this construction in terms of stochastic processes
related to Cuntz algebras.

\smallskip
\subsection{Potentials with the Keane condition}

On the Cantor set $\Lambda=\cW_\fA^\omega$, we consider $\R_+$-valued
potentials $W_\beta$ satisfying the {\em Keane condition}:
\begin{equation}\label{Keane1}
 \sum_{a\in \fA} W_\beta(ax)=1, \ \ \ \forall x\in  \cW_\fA^\omega. 
\end{equation}

\smallskip

This condition has a direct interpretation in terms of Ruelle transfer operators
$$ \cR_{\sigma, W,\beta} \, f(x)= \sum_{\sigma(y)=x} W_\beta(y) f(y) =
\sum_{a\in \fA} W_\beta(ax)\, f(ax). $$
Namely, the Keane condition implies that $f(x)\equiv 1$ is fixed point of $\cR_{\sigma, W,\beta}$.

\smallskip

The choice of a Keane potential $W_\beta$ on $\Lambda$ gives rise to a 
multifractal measure on $\Lambda$ obtained as a stochastic process as follows.
Choose a base point $x_0\in \cW_\fA^\omega$, and define the measure $\mu_{W,\beta,x_0}$
by setting
\begin{equation}\label{muWx0}
 \mu_{W,\beta,x_0}(\Lambda(w))= W_\beta(a_1x_0) W_\beta(a_2 a_1 x_0)\cdots
W_\beta(a_m \cdots a_2 a_1x_0) 
\end{equation}
for $w=a_0 \cdots a_m \in \cW_{\fA,m}$ and
$$ \Lambda(w)=\{ a \in \cW_\fA^\omega\,|\, a_0 \cdots a_m=w \}. $$
The Keane condition ensures that \eqref{muWx0} indeed defines a measure,
see \cite{DuJo}, \cite{MaPa}.

\smallskip

We focus in particular on two examples of potentials $W_\beta$, already
considered in \cite{ManMar} in relation to coding theory. The first example
gives a stochastic process governed by a Bernoulli measure and the second
one by a Markov measure.

\begin{ex}\label{W1ex}{\rm For $x=x_1 x_2 x_3 \cdots x_n \cdots \in \Lambda$, set
$W_\beta(x)=e^{-\beta \lambda_{x_1}}$, 
where the weights $\{ \lambda_a \}_{a\in \fA}$ satisfy
$$ \sum_{a\in \fA} e^{-\beta \lambda_a} =1. $$
Then the multifractal measure on $\Lambda$ is given by
$$ \mu_{W,\beta,x}(\Lambda(w))= \prod_{j=0}^m e^{-\beta \lambda_{w_j}}, $$
for $w=w_1 w_2 \cdots w_m \in \cW_{\fA,m}$, and $w_0=x_1$.}
\end{ex}

\begin{ex}\label{W2ex}{\rm
For $x=x_1 x_2 x_3 \cdots x_n \cdots \in \Lambda$, set
$W_\beta(x)=e^{-\beta \lambda_{x_1 x_2}}$, 
where the matrix $(\lambda_{ab})_{a,b\in \fA}$ satisfies the stochastic condition
$$  \sum_{a\in \fA} e^{-\beta \lambda_{a b}} =1 , \ \ \forall b\in \fA . $$
Then the measure is given by
$$ \mu_{W,\beta,x}(\Lambda(w))= e^{-\beta \lambda_{w_m w_{m-1}}}
\cdots e^{-\beta \lambda_{w_2 w_1}} e^{-\beta \lambda_{w_1 x_1}}. $$ }
\end{ex}

We will see how to adapt the second example to match the required properties
for a stochastic process on the Eternal Symmetree.

\smallskip
\subsection{Cuntz algebras}

Stochastic processes of the type \eqref{muWx0} were considered
as a source of wavelet constructions in \cite{DuJo}, and related to
representations of Cuntz algebras.

\smallskip

Given a finite set $\fA$, the Cuntz algebra $\cO_\fA$ is the 
universal $C^*$-algebra generated by {\em isometries} $S_a$ with $a\in \fA$
with the relation
$$ \sum_{a\in \fA} S_a S_a^* = 1 $$
and $S^*_a S_b =\delta_{a,b}$.

\smallskip

The maximal abelian subalgebra of $\cO_\fA$ is generated by
the projections $P_w =S_w S_w^*$ with $w\in \cW_{\fA}^\star$.
It is isomorphic to the $C^*$-algebra of continuous
functions $C(\Lambda_{\fA})$ on a Cantor set $\Lambda_{\fA}=\cW_{\fA}^\omega$. 
For more details on the properties of Cuntz algebras, we refer the reader to \cite{Cu}.

\smallskip

We will discuss in detail the relation of Cuntz algebras to the multifractal 
measures $\Lambda_{\fA}$ and stochastic processes arising from potentials 
with the Keane conditions in \S \ref{KMSsec} below, in the more general
case of Cuntz--Krieger algebras.

\smallskip
\subsection{Multifractal measures and stochastics on the Eternal Symmetree}

We now show that a stochastic process $P_a = e^{S_a/2} \Phi_a$
satisfying \eqref{Phiproc} on the Eternal Symmetree without pruning
can be obtained as the multifractal measure determined by a particular 
choice of Keane potential on the Cantor set $\Lambda_{\fA}$.

\begin{prop}\label{StochWS}
Let $\cS=(\cS_{ab})$ be a symmetric positive stochastic matrix with Perron--Frobenius eigenvalue
$\lambda_\cS=1$ and positive (left) Perron--Frobenius eigenvector $v_{\cS}$, 
normalized by $\sum_{a\in \fA} v_{\cS,a} =1$. Then
\begin{equation}\label{WSv}
 W(x)=v_{\cS,x_1}\, \cS_{x_1,x_2}\, v_{\cS, x_2}^{-1} 
\end{equation} 
defines a potential on $\Lambda_{\fA}$ satisfying the Keane condition.
The multifractal measure $\mu_{W,x}$ on $\Lambda_{\fA}$ determined
by $W$ and the choice of a base point $x$ defines a stochastic process
$\Phi_{x,a}(m)$ on the Eternal Symmetree without pruning, 
satisfying \eqref{Phiproc}.
\end{prop}

\proof The potential \eqref{WSv} satisfies the Keane condition \eqref{Keane1}, 
since we have
$$  \sum_{a\in \fA} W(ax)= \sum_{a\in \fA} 
v_{\cS,a}\, \cS_{a,x_1}\, v_{\cS, x_1}^{-1} = v_{\cS,x_1}\,  v_{\cS, x_1}^{-1} =1, $$
by the Perron--Frobenius condition. We interpret the coordinates of the 
Perron--Frobenius eigenvector $v_{\cS}$ in terms of the entropies of the 
Symmetree model, by setting $v_{\cS,a}=e^{-S_a/2}$. 
The resulting multifractal measure $\mu_{W,x}$ is then 
given by
$$  \mu_{W,x}(\Lambda(w)) =  e^{(S_{x_1}- S_{w_m})/2} 
\cS_{w_m w_{m-1}} \cdots \cS_{w_2 w_1} \cS_{w_1,x_1} ,$$
for a choice of an endpoint $x\in \Lambda_{\fA}$ and of 
a finite word $w=w_m w_{m-1}\cdots w_1$ in $\cW^\star_\fA$. 
Setting $\Phi_{x,a}(m+1)= e^{S_a/2} \sum_w \mu_{W,x}(\Lambda(aw))$,
we obtain
$$  \mu_{W,x}(\Lambda(aw)) = e^{(S_{x_1}- S_{a})/2} 
\cS_{a w_m} \cS_{w_m w_{m-1}} \cdots \cS_{w_2 w_1} \cS_{w_1,x_1} .$$
This satisfies
$$ \sum_w \mu_{W,x}(\Lambda(aw)) = e^{(S_{x_1}- S_{a})/2}  
\sum_w \cS_{a w_m} \cS_{w_m w_{m-1}} \cdots \cS_{w_2 w_1} \cS_{w_1,x_1}
= e^{(S_{x_1}- S_{a})/2}  \cS^m_{a,x_1}. $$ 
Thus, we obtain that $\Phi_{x,a}(m+1)= e^{S_a/2} \sum_w \mu_{W,x}(\Lambda(aw))$
satisfies \eqref{Phiproc}, since
$$ \Phi_{x,a}(m+1)= e^{S_{x_1}/2}  \cS^m_{a,x_1}=e^{S_{x_1}/2} \sum_b \cS_{a,b} 
\cS^{m-1}_{b,x_1} = \sum_b \cS_{a,b} \Phi_{x,b}(m). $$
\endproof

\begin{rem}\label{intLambda}
{\rm The dependence on the choice of a basepoint $x\in \Lambda_\fA$
of the random process $\Phi_{x,a}(m)$ constructed in Proposition 
\ref{StochWS} can be averaged out by setting
\begin{equation}\label{intPhi}
\Phi_a(m) = \int_{\Lambda_{\fA}} \Phi_{x,a}(m) \, d\mu(x),
\end{equation}
with $\mu$ the normalized Hausdorff measure on the Cantor 
set $\Lambda_{\fA}$. This still satisfies $\Phi (m+1) = \cS \, \Phi(m)$.
}
\end{rem}

\smallskip

It is convenient for our purposes, to rewrite the potential $W$ of \eqref{WSv}
and the resulting stochastic process in terms of an auxiliary parameter $\beta$,
which will play the role of a thermodynamic parameter when we reintepret the
construction as arising from a quantum statistical mechanical system. 

\begin{cor}\label{WandWbeta}
For a given $\beta>0$, the potential \eqref{WSv} can be
obtained as a particular case of Example \ref{W2ex} with
\begin{equation}\label{Westree}
 W_\beta(x) = e^{-\beta \lambda_{x_1 x_2}} = 
e^{-\beta \lambda_{x_1}} \, \cS_{x_1 x_2}\, e^{\beta \lambda_{x_2}} ,
\end{equation}
with $\cS=(S_{ab})_{a,b\in \fA}$ a symmetric positive stochastic matrix
as in Proposition \ref{StochWS} and with the weights $\lambda_a$ chosen so that 
\begin{equation}\label{vSbetalambda}
 e^{-\beta \lambda_a}= (v_S)_a = e^{-S_a/2},
\end{equation}
where, as above, $v_\cS$ is  the (left) Perron-Frobenius eigenvector, normalized
by  
\begin{equation}\label{PFnorm}
\sum_{a\in \fA} e^{-\beta \lambda_a}=1 .
\end{equation}
\end{cor}

\smallskip

The resulting multifractal measure, in the notation of 
Corollary \ref{WandWbeta}, is given by
\begin{equation}\label{muSproc}
 \mu_{W,\beta,x}(\Lambda(w)) = 
 e^{-\beta \lambda_{w_m w_{m-1}}} \cdots
e^{-\beta \lambda_{w_2 w_1}}e^{-\beta \lambda_{w_1 x_1}}  
 =  e^{\beta (\lambda_{x_1}- \lambda_{w_m})} 
\cS_{w_m w_{m-1}} \cdots \cS_{w_2 w_1} \cS_{w_1,x_1} .
\end{equation}

\medskip

\begin{cor}\label{equilsol}
Let $e^{-\tilde S_a/2}=\tilde v_{\cS,a}$, with $\tilde v_\cS$ a (right) Perron-Frobenius 
eigenvector for $\cS$.
Setting $P_a(m)= e^{-\tilde S_a/2} /\cN$ with $\cN=\sum_a e^{-\tilde S_a/2}$ gives
a stationary process satisfying $P_a(m+1)=\sum_b S_{ab} P_b(m)$ and
$P_a(m+1)=P_a(m)$.
\end{cor}

\proof Since $e^{-\tilde S_a/2}=\tilde v_{\cS,a}$ 
is a (right) Perron--Frobenius eigenvector of $\cS$ with $\lambda_{\cS}=1$,
we have $P_a(m+1)=\sum_b S_{ab} P_b (m) = 
\sum_b S_{ab} e^{-\tilde S_a/2} /\cN= e^{-\tilde S_b/2}/\cN=P_a(m)$.
\endproof

\smallskip
\subsection{Terminal vacua via Cuntz--Krieger algebras}

Let $\fA$ be a finite set and let $A=(A_{ab})_{a,b\in \fA}$ be a matrix with
entries $A_{a,b}\in \{ 0,1 \}$. We use the matrix $A$ as a way of pruning
the Bethe tree. The analog of the Cuntz algebra in the case of a subshift 
of finite type with admissibility condition given by $A$ is given by the
Cuntz--Krieger algebra $\cO_{\fA,A}$, see \cite{CK}.

\smallskip

The Cuntz--Krieger algebra $\cO_{\fA,A}$ is the universal $C^*$-algebra generated by 
{\em partial isometries} $S_a$ with $a\in \fA$, with relations
$$ S_a^* S_a = \sum_b A_{ab}\, S_b S_b^*, $$
$$ \sum_{a\in \fA} S_a S_a^* =1 . $$
The Cuntz algebras recalled above correspond to the special case where
the matrix $A$ has all entries equal to one, that is, to the unpruned case.

\smallskip

The maximal abelian subalgebra of  $\cO_{\fA,A}$ is generated by the projections
$S_w S_w^*$, with $w\in \cW^\star_{\fA,A}$ words with admissibility condition
$A_{w_k, w_{k+1}}=1$, in the language $\cL_{A}$. It is isomorphic to the
$C^*$-algebra $C(\Lambda_{\fA,A})$ of continuous functions on the Cantor
set $\Lambda_{\fA,A}$. This is the set  $\cL^\omega_A$ of infinite admissible words,
or equivalently the  endpoints at infinity of the {\em pruned tree} in the Eternal 
Simmetree model.

\smallskip
\subsection{Random processes with Keane potentials}\label{CKpotSec}

In the case of a subshift of finite type with matrix $A$, the Keane condition
for a potential $W_\beta$ on $\Lambda_{\fA,A}$ is given by
\begin{equation}\label{Keane2}
 \sum_{a\in \fA} A_{a x_1} \, W_\beta(\sigma_a(x))=1, 
\end{equation}
which is equivalent to the condition $\sum_{y: \sigma(y)=x} W_\beta(y)=1$,
written in terms of the shift map $\sigma: \Lambda_{\fA,A}\to \Lambda_{\fA,A}$.

\smallskip

The associated random process, defining a multifractal measure on 
$\Lambda_{\fA,A}$, is given by
$$ \mu_{W,\beta,x_0}(\Lambda_{\fA,A^t}(w))= A_{w_1 x_1} W_\beta(\sigma_{w_1}(x))
\cdots W_\beta(\sigma_{w_n}\cdots \sigma_{w_1}(x)). $$
This satisfies
$$ \mu_{W,\beta,x_0}(\Lambda_{\fA,A^t}(w))= \sum_b A^t_{w_k b} \,
 \mu_{W,\beta,x_0}(\Lambda_{\fA,A^t}(wb)). $$

\smallskip

This random process is related to fixed points of the Ruelle
transfer operator (see \cite{MaPa}), 
\begin{equation}\label{RuelleCK}
 \cR_{\sigma,W} f (x)= \sum_{y: \sigma(y)=x} W_\beta(y) f(y) =
\sum_{a\in \fA} A_{a x_1} \, W_\beta(\sigma_a(x)) f(\sigma_a(x)) . 
\end{equation}
The Ruelle transfer operator \eqref{RuelleCK}, with a
potential $W_\beta :\Lambda_{\fA,A} \to \R^*_+$, can be written equivalently
in terms of elements in the Cuntz--Krieger algebra $\cO_{\fA,A}$ as
$$ \cR_{\sigma,W} f = \sum_{a\in \fA} S_a^* \, W_\beta \, f \, S_a. $$

\smallskip
\subsection{Quantum Statistical Mechanics, KMS states, and self-similar measures}\label{KMSsec}

In order to explain the relation between the multifractal measures on $\Lambda_\fA$
and $\Lambda_{\fA,A}$ constructed via potentials with the Keane condition and the
operator algebras of Cuntz and Cuntz--Krieger type, we need to recall some preliminary
notions about operator algebra based Quantum Statistical Mechanics. We refer
the reader to \cite{BrRo} for a detailed and comprehensive introduction to the subject.

\smallskip

The basic data of a quantum statistical mechanical system consist of:
\begin{itemize}
\item A unital $C^*$-algebra $\cA$ of {\em observables};
\item A {\em time evolution}, given by a one-parameter family of automorphisms 
$\sigma: \R \to {\rm Aut}(\cA)$;
\item States, given by continuous linear functionals $\varphi: \cA \to \C$ with a
positivity condition $\varphi(a^* a)\geq 0$ and normalized to $\varphi(1)=1$;
\item Equilibrium states, satisfying $\varphi(\sigma_t(a))=\varphi(a)$.
\end{itemize}

\smallskip

In particular, an important class of equilibrium states is given by 
{\em KMS states} at inverse temperature $\beta$: these are states that
satisfy the condition 
$$ \varphi_\beta(ab) = \varphi_\beta(b\sigma_{i\beta}(a) ) $$
for all $a,b$ in a dense subalgebra of ``analytic elements" (that is,
elements for which the time evolution $\sigma_t$ admits an 
analytic continuation to $\sigma_z$, with $z$ in a strip of height $\beta$
in the complex upper half plane. 

\smallskip

A typical example of KMS states is given by Gibbs states, of the form
$$ \varphi_\beta(a)= \frac{{\rm Tr}(\pi(a) e^{-\beta H})}{{\rm Tr}(e^{-\beta H})} $$
for $\pi$ a Hilbert space representation of the algebra $\cA$ and
$H$ the infinitesimal generator of the time evolution in the representation,
$\pi(\sigma_t(a))=e^{it H} \pi(a) e^{-itH}$. Gibbs states are well defined
only under the condition that ${\rm Tr}(e^{-\beta H})<\infty$, while KMS states
exist in greater generality. Indeed, the KMS states that we will be considering
on Cuntz and Cuntz--Krieger algebras, related to multifractal measures,
are {\em not} of Gibbs form.

\smallskip

The first example of self-similar measure on $\Lambda_{\fA,A}$ that can be obtained
from Quantum Statistical Mechanics on the Cuntz--Krieger algebra $\cO_{\fA,A}$ is
determined by the Perron--Frobenius theory of the matrix $A$, as in \cite{MaPa}.
Setting $\mu(\Lambda_{\fA,A}(w)) = \lambda_A^{-k} (v_A)_{w_k}$, for
$w=w_1\cdots w_k \in \cW_{\fA,A}^\star$, with $v_A=(v_A)_a$ the
Perron--Frobenius eigenvector of $A$, determines a measure satisfying the
self-similarity condition 
$$ \mu = \lambda_A^{-1} \, \sum_{a\in \fA} \mu \circ \sigma_a^{-1}, $$
where $\lambda_A$ is the Perron--Frobenius eigenvalue of $A$. 
This is a fractal measure with Hausdorff dimension 
$\delta_A= \log(\lambda_A)/\log(\#\fA)=\dim_H(\Lambda_{\fA,A})$. 

\smallskip

This fractal measure can be obtained (see \cite{MaPa}) by considering
the time evolution on the Cuntz--Krieger algebra $\cO_{\fA,A}$ determined
by setting $\sigma_t(S_a)=q^{it}\, S_a$ with $q=\# \fA$. This time
evolution has a unique temperature at which KMS states exist, which
is equal to the Hausdorff dimension, $\beta = \delta_A$. At this
temperature there is a unique KMS state, given by
$$ \varphi(S_w S_v^*)= \left\{ \begin{array}{ll} 0 & v\neq w \\
\mu(\Lambda_{\fA,A}(w)) & v=w \in \cW_{\fA,A}^\star
\end{array}\right. $$
which determines and is in turned determined by the self-similar
measure on  $\Lambda_{\fA,A}$.

\smallskip

A more general result relating quantum statistical mechanics
on Cuntz--Krieger algebras to multifractal measures was 
obtained in \cite{KeStSt}. 

\smallskip

The following result is well known from the work of \cite{KeStSt}.
We report it here for convenience.

\begin{lem}\label{KMSnu}
Consider a potential $W:\Lambda_{\fA,A} \to \R^*_+$, with
$W_\beta(x)=W(x)^{-\beta}$. Consider the time evolution on $\cO_{\fA,A}$ 
defined by 
\begin{equation}\label{sigmatWbeta}
\sigma_t(S_a)=W^{it}\, S_a.
\end{equation}
Then KMS$_\beta$ states $\varphi_\beta$ 
for $(\cO_{\fA,A}, \sigma_t)$ determine multifractal measures 
on $\Lambda_{\fA,A}$ that are fixed by the dual Perron--Frobenius
operator $\cR^*_{\sigma,W,\beta} \nu_{W,\beta} =\nu_{W,\beta}$.
\end{lem}

\proof As observed in Fact 8 of \cite{KeStSt}, by gauge
invariance, a KMS$_\beta$ state $\varphi_\beta$ 
for this time evolution satisfies 
$\varphi_\beta(S_w S_{w'}^*)=0$ for all
$w\neq w' \in  \cW_{\fA,A,k}$. Moreover, by Fact 7
of \cite{KeStSt}, its restriction to the 
subalgebra $C(\Lambda_{\fA,A})$ determines
a measure, which is a fixed point of the dual Perron--Frobenius
operator $\cR^*_{\sigma,W,\beta} \nu_{W,\beta} =\nu_{W,\beta}$.
\endproof

Under the assumption that $W=e^H$ with $H\geq 0$ and with $\mu(\{ H=0 \})=0$,
there is in fact a bijection between KMS$_\beta$ states and fixed points of
$\cR^*_{\sigma,W,\beta}$, see Fact 9 of \cite{KeStSt}. 

\smallskip

We look in particular at the cases described in Examples \ref{W1ex} and \ref{W2ex}.
In order to adapt Example \ref{W1ex} from the Cuntz to the Cuntz--Krieger case,
we need to assume that the potential $W_\beta(x)=e^{-\beta x_1}$ satisfies the
Keane condition \eqref{Keane2} instead of \eqref{Keane1}. This means requiring
that $\sum_a A_{ab} e^{-\beta \lambda_a} =1$ for all $b\in \fA$. This is possible
if $A$ is invertible and the weights $\lambda_a$ are chosen (depending on $\beta$)
so that the vector with entries $(e^{-\beta \lambda_a})_{a\in \fA}$ is $A^{-1}{\bf 1}$,
where ${\bf 1}$ is the vector with all entries equal to one.
The following shows how one can realize a stochastic process given by a Bernoulli
measure as a KMS state.

\smallskip

We make here some simplifying assumptions on the matrix $A$, though a similar
statement can be formulated more generally ({\em mutatis mutandis}).

\begin{lem}\label{ex1KMSnu}
Assume that $A$ is invertible and symmetric and 
consider a potential $W:\Lambda_{\fA,A} \to \R^*_+$ as in Example \ref{W1ex}, 
satisfying the Keane condition \eqref{Keane2}. Consider the
time evolution \eqref{sigmatWbeta} on $\cO_{\fA,A}$. Then a
KMS$_\beta$ state $\varphi_\beta$ is obtained by considering the measure
$$ \nu_{W,\beta}(\Lambda_{\fA,A}(w)) = \prod_{j=1}^k e^{-\beta \lambda_{w_j}} . $$
\end{lem}

\proof A KMS$_\beta$ $\varphi_\beta$ determines a measure $\nu_{W,\beta}$
as in Lemma \ref{KMSnu}.  Using the KMS-property
we see that the measure $\nu_W$ satisfies
$$ \nu_{W,\beta}(\Lambda_{\fA,A}(w))= \varphi_\beta( S_w S_w^*)=
\varphi_\beta( S_{w_2}\cdots S_{w_k} S_w^* \sigma_{i\beta}(S_{w_1})) . $$
We have $S_a^* W = (W\circ \sigma_a)\, \chi_{D_a}  \, S_a^*$, where $D_a$ is the
domain of the partial inverse $\sigma_a$ of the shift map, namely $D_a=\{ x\,|\,
A_{ax_1}=1\}$. Thus, we can write the above as
$$ \varphi_\beta(S_{w_2}\cdots S_{w_k} S_{w_k}^* \cdots S_{w_2}^* \,\, 
W_\beta\circ \sigma_{w_1}\, \, P_{w_1}) = \varphi_\beta(W_\beta\circ \sigma_w \,\, W_\beta\circ \sigma_{w_k \cdots w_2} \cdots W_\beta\circ \sigma_{w_k} \,\, P_w), $$
where $P_w=S_w^* S_w$. We can write the above in the form 
$$ \nu_{W,\beta}(\Lambda_{\fA,A}(w))= \int_{\Lambda_{\fA,A}} 
W_\beta\circ \sigma_w \, \, W_\beta\circ \sigma_{w_k \cdots w_2} \cdots W_\beta\circ \sigma_{w_k} \,\, 
\chi_{D_w} \, d\nu $$
$$ =  \int_{\Lambda_{\fA,A}} A_{w_k x_1}\, 
W_\beta(w_1 \cdots w_k x) \,\, W_\beta(w_2 \cdots w_k x) \cdots 
W_\beta(w_k x) \,\, d\nu(x), $$
where $\nu$ is a probability measure on $\Lambda_{\fA,A}(w)$.
In the case of a locally constant potential $W_\beta(x)=e^{-\beta \lambda_{x_1}}$
that only depends on the first digit of $x\in \Lambda_{\fA,A}$ we have
$$ \nu_{W,\beta}(\Lambda_{\fA,A}(w))= e^{-\beta \lambda_{w_1}}\cdots e^{-\beta \lambda_{w_k}} 
\int_{\Lambda_{\fA,A}} A_{w_k x_1} d\nu(x) = 
e^{-\beta \lambda_{w_1}}\cdots e^{-\beta \lambda_{w_k}} 
\sum_{a\in \fA} A_{w_k a} \, \nu(\Lambda_{\fA,A}(a)). $$
If $\nu(\Lambda_{\fA,A}(a))=e^{-\beta \lambda_a}$ the Keane condition with $A=A^t$ gives
that the sum is equal to one, hence $\nu_{W,\beta}(\Lambda_{\fA,A}(w))$ is as stated.
\endproof

\medskip

We now consider the more interesting case of stochastic processes governed
by a Markov measure as in Example \ref{W2ex}. In the Cuntz--Krieger case,
this means that we consider a locally constant 
potential $W_\beta(x)=e^{-\beta \lambda_{x_1 x_2}}$, which depends on the
first two digits of $x\in \Lambda_{\fA,A}$ satisfying the Keane condition \eqref{Keane2},
\begin{equation}\label{Keane3}
 \sum_{a\in \fA} A_{a b} \, e^{-\beta \lambda_{a b}} =1, \ \ \  \forall b\in \fA. 
\end{equation} 
This condition means that the matrix $T=(T_{ab})_{a,b\in \fA}$ with
$T_{ab}=A_{a b} \, e^{-\beta \lambda_{a b}}$ is a stochastic matrix.
Recall that a non-negative matrix $M$ is irreducible if for every pair of indices $i,j$
there is an $m>0$ such that $M^m_{ij}>0$.

\smallskip

\begin{lem}\label{PFofT}
Assume that the matrix $T$ defined above is irreducible. Then $T$ has Perron--Frobenius
eigenvalue $\lambda_T=1$, with one-dimensional eigenspace and 
with a (right) Perron--Frobenius eigenvector $v_T$ with positive 
components, $v_{T,a}>0$, for all $a\in \fA$. 
\end{lem}

\proof The dimension of the eigenspace and the positivity of the 
Perron--Frobenius eigenvector result from the Perron--Frobenius
theorem for irreducible non-negative matrices. The fact that the
eigenvalue $\lambda_T=1$ follows from the estimate
$$ \min_b \sum_a   A_{a b} \, e^{-\beta \lambda_{a b}} \leq \lambda_T \leq
\max_b \sum_a   A_{a b} \, e^{-\beta \lambda_{a b}} $$
and the Keane condition.
\endproof

\smallskip

We can then obtain a stochastic process of Markov type as in Example
\ref{W2ex} as a KMS state in the following way.

\begin{prop}\label{ex2KMSnu}
Consider a locally constant potential $W:\Lambda_{\fA,A} \to \R^*_+$
of the form $W(x)=e^{\lambda_{x_1,x_2}}$, with $W_\beta=W^{-\beta}$
satisfying the Keane condition \eqref{Keane3}.
Assume that $T=(A_{a b} \, e^{-\beta \lambda_{a b}})$ is irreducible. 
Consider the time evolution \eqref{sigmatWbeta} on $\cO_{\fA,A}$. Then a
KMS$_\beta$ state $\varphi_\beta$ is obtained by considering the measure
\begin{equation}\label{nuvT}
 \nu_{W,\beta}(\Lambda_{\fA,A}(w)) = \prod_{j=1}^{k-1} e^{-\beta \lambda_{w_j, w_{j+1}}}
\, v_{T,w_k} , 
\end{equation}
where $v_T=(v_{T,a})_{a\in \fA}$ is the  positive (right) Perron--Frobenius eigenvector of $T$,
normalized by the condition $\sum_{a\in \fA} v_{T,a}=1$.
\end{prop}

\proof We proceed as in Lemma \ref{ex1KMSnu}. The values of $\varphi_\beta$
on the elements $S_w S_w^*$ are given by a measure
$$  \nu_{W,\beta}(\Lambda_{\fA,A}(w))=  \int_{\Lambda_{\fA,A}} A_{w_k x_1}\, 
W_\beta(w_1 \cdots w_k x) \,\, W_\beta(w_2 \cdots w_k x) \cdots 
W_\beta(w_k x) \,\, d\nu(x) $$
$$ =  e^{-\beta \lambda_{w_1 w_2}}  \cdots e^{-\beta \lambda_{w_{k-1} w_k}} \, 
\int_{\Lambda_{\fA,A}} A_{w_k x_1}\, e^{-\beta \lambda_{w_k x_1}}\, d\nu(x) $$ $$ =
e^{-\beta \lambda_{w_1 w_2}}  \cdots e^{-\beta \lambda_{w_{k-1} w_k}} \, 
\sum_{a\in \fA} A_{w_k a} \, e^{-\beta \lambda_{w_k a}}\, \nu(\Lambda_{\fA,A}(a)). $$
Setting $\nu(\Lambda_{\fA,A}(a))=v_{T,a}$ gives a probability measure on $\Lambda_{\fA,A}$
and we get
$$ \nu_{W,\beta}(\Lambda_{\fA,A}(w))= e^{-\beta \lambda_{w_1 w_2}} 
 \cdots e^{-\beta \lambda_{w_{k-1} w_k}} \, v_{T,w_k}, $$
 since by Lemma \ref{PFofT} the Perron--Frobenius
eigenvalue $\lambda_T=1$.
\endproof

\medskip
\subsection{Random processes with terminal vacua and multifractal measures}

The construction of the random process on the pruned Eternal Symmetree
is analogous to the non-pruned case, but taking into account the presence
of the pruning, through the admissibility matrix $A=(A_{ab})$. This means
that we use the type of random process associated to Cuntz--Krieger
algebras, as shown in \S \ref{CKpotSec} above. More precisely, we show
here that the stochastic process associated to the KMS state on the
Cuntz--Krieger algebra in Proposition \ref{ex2KMSnu} above can be
used to obtain the type of stochastic process considered in \cite{HSSS}
in the Eternal Symmetree model with pruning.

\smallskip

As in the unpruned case, let $\cS=(\cS_{ab})$ be a positive stochastic
matrix, and set $\tilde\cS=A \cS$. This is a non-negative matrix with
Perron--Frobenius $\lambda_{\tilde\cS} <1$ and Perron--Frobenius 
eigenvector $v_{\tilde\cS}=(v_{\tilde\cS,a})_{a\in \fA}$, with $v_{\tilde\cS,a}>0$.
\smallskip

As in Proposition \ref{ex2KMSnu}, consider a potential 
$W_\beta(x)=e^{-\beta \lambda_{x_1 x_2}}$ satisfying the
Keane condition \eqref{Keane3}. 

\smallskip

\begin{prop}\label{PFTtildeS}
Let $\beta$ and $\{ \lambda_a \}_{a\in \fA}$ be chosen so that 
$e^{-\beta \lambda_a}=v_{\tilde \cS,a}$ are the components of the normalized (right) 
 Perron--Frobenius eigenvector of  $\tilde\cS=A \cS$, with eigenvalue $\lambda_{\tilde\cS}$. Let 
 \begin{equation}\label{WbetatildeS}
 W_\beta(x)= e^{-\beta \lambda_{x_1 x_2}}= \frac{1}{\lambda_{\tilde\cS}}\, e^{-\beta \lambda_a}\,
\cS_{ab} \, e^{\beta \lambda_b}. 
\end{equation}
Then $W_\beta$ satisfies the Keane condition \eqref{Keane3}.
Let $T_{ab} = A_{ab} e^{-\beta \lambda_{a b}}$ with Perron--Frobenius eigenvector $v_T$ with
eigenvalue $\lambda_T=1$. The components of $v_T$ satisfy $v_{T,a}=q^{-1}$ for all $a\in \fA$,
with $q=\# \fA$.
\end{prop}

\proof The potential $W_\beta$ of \eqref{WbetatildeS} satisfies the Keane condition since
$$ \frac{e^{\beta \lambda_b}}{\lambda_{\tilde\cS}}\,  \sum_a A_{ab} \cS_{ab}  
e^{-\beta \lambda_a} =1, $$
since $e^{-\beta \lambda_a}=v_{\tilde S,a}$. The Perron--Frobenius condition
$$ \sum_a A_{ab} e^{-\beta \lambda_{ab}} \, v_{T,a} = v_{T,b} $$
then implies
$$ \frac{e^{\beta \lambda_b}}{\lambda_{\tilde\cS}}\, \sum_a \tilde\cS_{ab} \, e^{-\beta\lambda_a}
\, v_{T,a} = v_{T,b} , $$
hence $e^{-\beta \lambda_a}\, v_{T,a}= \alpha \, v_{\tilde \cS,a}$. Since
$e^{-\beta \lambda_a}=v_{\tilde \cS,a}$, we have $v_{T,a}=\alpha$, the uniform
measure, with $\alpha$ fixed by the normalization $\sum_a v_{T,a}=1$ to be $\alpha=q^{-1}$,
with $q=\# \fA$. 
\endproof

\smallskip

\begin{cor}\label{corSnuWbeta}
The measure $\nu_{W,\beta}$ associated to the potential \eqref{WbetatildeS} is given by
\begin{equation}\label{SnuWbeta}
\nu_{W,\beta}(\Lambda_{\fA,A}(w))= \frac{e^{\beta (\lambda_{w_k}-\lambda_{w_1})}}
{q\, \lambda_{\tilde \cS}^{k-1}} \, \cS_{w_1 w_2} \cdots \cS_{w_{k-1} w_k}.
\end{equation}
\end{cor}

\proof This follows directly by \eqref{nuvT} and Proposition \ref{PFTtildeS}.
\endproof

\smallskip

Notice that here, unlike in \cite{HSSS}, we are maintaining the normalization
of the measure, by maintaining the Keane condition on the potential, which
results in dividing by an increasingly large power of the Perron--Frobenius
eigenvalue $\lambda_{\tilde\cS}$ in \eqref{SnuWbeta}. This is equivalent
to the observation of  \cite{HSSS}, \cite{Suss} that, in the presence of
terminal vacua, the eternal inflation is concentrated on a fractal set of
increasingly small volume (scaling by a power of $\lambda_{\tilde\cS}$),
when measured with respect to the original stochastic process of the
unpruned tree. In the process \eqref{SnuWbeta} the volume remains
constant, but at the cost of a large dilation by powers of $\lambda_{\tilde S}^{-1}$.

\begin{cor}\label{Philambda}
For entropies $S_a$ satisfying $e^{-S_a/2}/\cN=e^{-\beta \lambda_a}$, with the 
normalization factor $\cN=\sum_a e^{-S_a/2}$, consider the process
$$ \Phi_a(m+1)=\lambda_{\tilde\cS}^m e^{S_a/2} \sum_{w\in \cW_{\fA,A,m}} 
\nu_{W,\beta}(\Lambda_{\fA,A}(aw)). $$
This determines a stochastic process on the pruned tree satisfying 
$\Phi_a(m+1)=\sum_b \cS_{ab} \Phi_b(m)$.
\end{cor}

\proof We have 
$$ \sum_{w\in \cW_{\fA,A,m}} \nu_{W,\beta}(\Lambda_{\fA,A}(wa))=
\frac{e^{-\beta \lambda_a}}{q\, \lambda_{\tilde S}^m}
\sum_w e^{\beta \lambda_{w_m}} \, \cS_{a w_1} \cS_{w_1 w_2} \cdots \cS_{w_{m-1} w_m}
= \frac{e^{-\beta \lambda_a}}{q\, \lambda_{\tilde S}^m} \sum_b \cS^{m-1}_{ab} e^{\beta \lambda_b}. $$
Thus, we obtain
$$ \lambda_{\tilde\cS}^m e^{S_a/2} \sum_{w\in \cW_{\fA,A,m}} 
\nu_{W,\beta}(\Lambda_{\fA,A}(aw)) =q^{-1} \sum_b \cS^{m-1}_{ab} e^{\beta \lambda_b} $$
On the other hand, we have
$$ \sum_b \cS_{ab} \Phi_b(m) = q^{-1} \sum_b \cS_{ab} \sum_c \cS^{m-2}_{bc} e^{\beta \lambda_c}
= q^{-1} \sum_c \cS^{m-1}_{ac} e^{\beta \lambda_c}. $$
\endproof

\medskip

\subsection{Multiverse fields and propagators}\label{multiverseSec}

In the Eternal Symmetree model, the random process described above
determines {\em multiverse fields} $\cO(x)$ with correlation functions
$\langle \cO(x), \cO(y) \rangle$ that depend on propagators $C_{a,b}(x,y)$
computed in \cite{HSSS} from the stochastic process. 

\smallskip

In the case without terminal vacua, the propagators on the
Eternal Symmetree are obtained as follows (\cite{HSSS}). 
Given $x,y \in \Lambda_\fA$, consider the rays starting at the
root vertex $v_0$ with ends $x$ and $y$, respectively, and
denote by $v_{x,y}$ the last common vertex between the two
rays. Let $w_{x,y}=w_1\cdots w_k$ be the finite word in 
$\cW^\star_{\fA}$ that labels the path starting at $v_0$ and 
ending at $v_{x,y}$. Then, for $a,b\in \fA$ and $x,y \in \Lambda_\fA$, 
the propagator $C_{a,b}(x,y)$ is given by
\begin{equation}\label{Cab}
 C_{a,b}(x,y) =\frac{1}{\cN} \sum_{c\in \fA}
e^{S_c} P_{a,c}(v_0,v_{x,y}) P_{b,c}(v_0,v_{x,y}), 
\end{equation}
with normalization factor $\cN=\sum_a e^{S_a}$, where
$P_{a,b}(v_0,v)$ is the probability, according to the
stochastic process on the tree, of going from a vacuum of
type $a$ to one of type $b$ along the path connecting $v_0$ to $v$.

\smallskip

\begin{prop}\label{CabS}
Using the random process of Proposition \ref{StochWS}, 
with $e^{\beta \lambda_a}=e^{S_a/2}$, we obtain
$$ C_{a,b}(x,y) =\frac{e^{(S_{x_1}+S_{y_1})/2}}{\cN}
\sum_{c\in \fA} \cS^{k-2}_{a,c} \cS_{c,x_1}
\cS^{k-2}_{b,c} \cS_{c,y_1}. $$
with $\cN=\sum_a e^{S_a}$, and with $w$ the word
labeling the path from $v_0$ to $v_{x,y}$.
\end{prop}

\proof We have
$$ C_{a,b}(x,y) =\frac{1}{\cN} \sum_{c\in \fA}
e^{S_c} \, \left(\sum_{w\in \cW_{\fA,A,k-2}} \mu_{W,x}(\Lambda(awc))\right) \, 
\left(\sum_{w\in \cW_{\fA,A,k-2}} 
\mu_{W,y}(\Lambda(bwc))\right) $$ $$ =\frac{1}{\cN} \sum_{c\in \fA} e^{S_c} \,
e^{(S_{x_1}-S_c)/2} \, \left(\sum_{w\in \cW_{\fA,A,k-2}} \cS_{a w_{k-1}} \cdots \cS_{w_2 c} \cS_{c x_1}\right)
e^{(S_{y_1}-S_c)/2} \, \left(\sum_{w\in \cW_{\fA,A,k-2}} \cS_{b w_{k-1}} \cdots \cS_{w_2 c} \cS_{c y_1}\right)
$$
$$ = \frac{e^{(S_{x_1}+S_{y_1})/2}}{\cN} \sum_{c\in \fA} \cS^{k-2}_{a,c} \cS_{c,x_1}
\cS^{k-2}_{b,c} \cS_{c,y_1}, $$
which gives the stated expression.
\endproof

In the case with terminal vacua we find the following. 

\begin{prop}\label{CabS}
Consider the case of the random process of Corollary \ref{corSnuWbeta}, with
$e^{\beta \lambda_a}=e^{S_a/2}$. We then have
$$ C_{a,b}(x,y) = \frac{e^{(S_a+S_b)/2}}{\cN q \lambda_{\tilde\cS}^{k-1}} \sum_{c\in \fA}
\cS^{k-2}_{c,b} \cS^{k-2}_{c,a}. $$
\end{prop}

\proof We have 
$$ C_{a,b}(x,y) = \frac{1}{\cN}  \sum_{c\in \fA} e^{S_c} (\sum_{w\in \cW_{\fA,A,k-2}}
\nu_{W,\beta}(\Lambda(cwa)))(\sum_{w\in \cW_{\fA,A,k-2}}
\nu_{W,\beta}(\Lambda(cwb))) $$
$$ = \frac{e^{(S_a+S_b)/2}}{\cN q \lambda_{\tilde\cS}^{k-1}}  \sum_{c\in \fA}
(\sum_{w\in \cW_{\fA,A,k-2}} \cS_{cw_2} \cdots \cS_{w_{k-1} a})
(\sum_{w\in \cW_{\fA,A,k-2}} \cS_{cw_2} \cdots \cS_{w_{k-1} b}), $$
which gives the statement.
\endproof

Again, here we have a power of $\lambda_{\tilde\cS}$ in the denominator
(instead of the numerator as in \cite{HSSS}) because we are measuring
volumes with respect to a measure that remains normalized on a fractal
that scales in size as a power of $\lambda_{\tilde\cS}$
(with respect to the original measure on the unpruned tree).

\smallskip

If one interprets the random processes obtained from Keane potentials
as a construction of wavelets on fractals, as in \cite{DuJo}, \cite{MaPa}, one
can interpret the propagators $C_{a,b}(x,y)$ of the 
Eternal Symmetree model as a measure of wavelet autocorrelation.

\section{Eternal inflation on Mumford curves}\label{MumSec}

We now consider a variation on the original idea of the Eternal Symmetree model
of eternal inflation constructed in \cite{HSSS}, where instead of working with a
subtree $\cT'$ of the Bruhat--Tits tree $\cT$ of (a finite extension of) $\Q_p$, we
consider quotients by actions of $p$-adic Schottky groups.

\smallskip

In the model we construct here, one still has infinite trees as in the original
Eternal Symmetree model, but these coexist with confined regions, where
the evolution induced by the flow on the covering Bruhat--Tits tree remains
confined behind a horizon. 

\smallskip

We show that the construction of a stochastic process on the Symmetree
can be extended to this case, after replacing the Cuntz--Krieger algebra
by a more general graph algebra with a time evolution, and the KMS state
with the more general notion of {\em graph weight} used in \cite{CMR},
related to modular index invariants.

\smallskip
\subsection{Mumford curves}

A $p$-adic Schottky group $\Gamma$ is a finitely generated, discrete, 
torsion-free subgroup of ${\rm PGL}_2(\bK)$, with $\bK$ a finite extension of $\Q_p$,
with the property that $\Gamma\simeq \Z^{\star g}$ (the free group on $g$-generators),
where all nontrivial elements are {\em hyperbolic}. The latter condition means that
all $\gamma\in \Gamma$, with $\gamma\neq 1$, have two fixed points $z^\pm(\gamma)$
located on the boundary $\P^1(\bK)$ of the Bruhat--Tits tree, on which $\Gamma$
acts by isometries. The {\em axis} $L(\gamma)$ is the geodesic in $\cT$ with
endpoints $z^\pm(\gamma)$. The {\em limit set} $\Lambda_\Gamma$ of the 
Schottky group $\Gamma$ is the closure in $\P^1(\bK)$ of the set of 
all the fixed points $\{ z^\pm(\gamma)\,|\, \gamma\neq 1 \in \Gamma \}$. 
The {\em domain of discontinuity} is the complement
$\Omega_\Gamma(\bK)=\P^1(\bK)\smallsetminus \Lambda_\Gamma$.
The quotient $X_\Gamma = \Omega_\Gamma/\Gamma$ is the $p$-adic
uniformization of an algebraic curve $X$ of genus $g$, a {\em Mumford curve}, see
\cite{Mum}.

\smallskip

We focus on the case of genus $g\geq 2$, where the limit set $\Lambda_\Gamma$
is a Cantor set, in contrast to the genus one case where it consists of just two
points $\{ 0, \infty \}$.

\smallskip

Let $\cT_\Gamma$ be the smallest subtree of the Bruhat--Tits tree $\cT$ of $\bK$
that contains all the axes $L(\gamma)$ of all $\gamma\neq 1$ in $\Gamma$.  This
satisfies $\partial \cT_\Gamma = \Lambda_\Gamma$. The action of $\Gamma$
preserves $\cT_\Gamma$ and the quotient $\cT_\Gamma/\Gamma$ is a finite graph.
In algebro-geometric terms, $G_\Gamma=\cT_\Gamma/\Gamma$ is the dual
graph of the closed fiber of the minimal smooth model of the curve $X$, \cite{Mum}.

\smallskip

The quotient $\cT/\Gamma$ of the action of $\Gamma$ on the full Bruhat--Tits tree
consists of a copy of the finite graph $G_\Gamma$ with infinite trees departing from
its vertices. The boundary at infinity is given by the algebraic points of the Mumford curve,
$$ \partial (\cT/\Gamma) = X_\Gamma(\bK) =\Omega_\Gamma(\bK) /\Gamma. $$

\smallskip

As shown in \cite{CMR}, there is always a choice of orientation on $\cT$ that
induces an orientation of the finite graph $G_\Gamma$ that extends to an
orientation of $\cT/\Gamma$ with outward pointing orientations on all the 
infinite trees attached to vertices of $G_\Gamma$.

\smallskip

\begin{center}
\begin{figure}
\includegraphics[scale=.55]{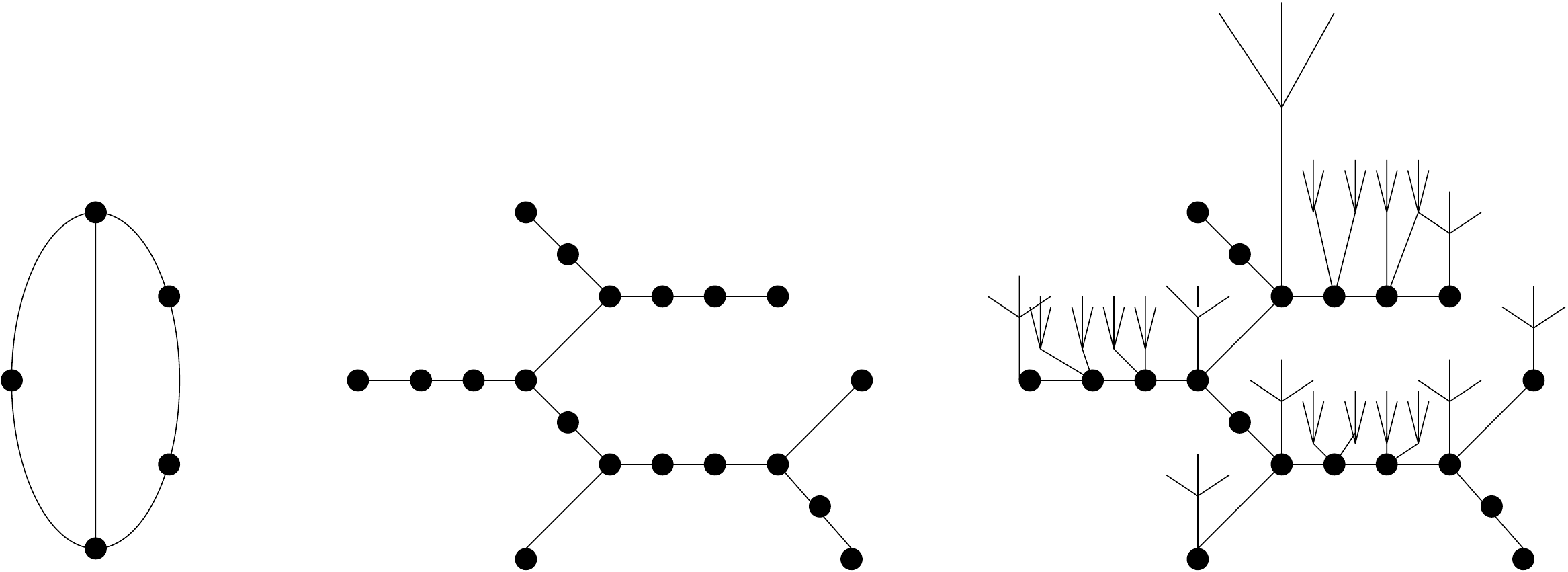}
\caption{A genus $g=2$ example: the finite graph $G_\Gamma$, the tree $\cT_\Gamma$,
and the rest of the Bruhat--Tits tree $\cT$ (from \cite{CMR}).}
\end{figure}
\end{center}

\smallskip
\subsection{Paths and horizons}

When we consider infinite paths in the Bruhat--Tits tree $\cT$ and their
image in the quotient $\cT/\Gamma$, we see that we obtain the following
types of behavior.
\begin{itemize}
\item Geodesics in $\cT$ with both endpoints in $\Omega_\Gamma$ give rise
to geodesics in $\cT/\Gamma$ with both endpoints on the curve at infinity $X_\Gamma$.
\item Geodesics in $\cT$ with one endpoint in $\Omega_\Gamma$ and one endpoint
in $\Lambda_\Gamma$ give rise to geodesics in $\cT/\Gamma$ which have a future 
(respectively, past) endpoint at infinity on the curve $X_\Gamma$ and that remain forever
confined in the past (respectively, future) inside the trapped region $G_\Gamma$.
\item Geodesics in $\cT$ with both endpoints on $\Lambda_\Gamma$ remain confined
in the trapped region $G_\Gamma$ in both the past and future direction.
\end{itemize}

An interpretation of Mumford curves as $p$-adic models of black holes,
in the context of the holography principle, was given in \cite{ManMar2}.

\smallskip
\subsection{Graph algebras and graph weights}

If we want to extend to the Mumford curves case the operator algebra 
approach to stochastic processes on the Symmetree developed in
the previous section, we need to replace the Cuntz and Cuntz--Krieger
algebras with more general graph algebras \cite{BPRS} based on Cuntz--Krieger
type relations, which we can use to model in operator theoretic terms
the graph $\cT/\Gamma$, as in \cite{CMR}.

\smallskip

Let $E=\cT/\Gamma$, with a fixed orientation of the finite graph 
$G_\Gamma=\cT_\Gamma/\Gamma$ and
with all the infinite trees oriented outward towards the boundary $X_\Gamma$.  
This is a directed graph with a countable number of vertices $E^0$
and of edges $E^1$. Let $s,t: E^1 \to E^0$ be the endpoint maps (source and
target) of the directed edges. A graph is {\em row-finite} if each vertex
has a finite number of outgoing edges and it is {\em locally finite} if 
each vertex has also a finite number of incoming edges. The graph $E=\cT/\Gamma$
satisfies these conditions, hence we apply the construction
of \cite{BPRS}, \cite{KPR} of graph $C^*$-algebras for row-finite and
locally finite graphs.

\smallskip

The graph $C^*$-algebra $C^*(E)$ is the universal $C^*$-algebra generated
by partial isometries $\{ S_e \}_{e\in E^1}$ satisfying the Cuntz--Krieger relations
\begin{equation}\label{CKgraph}
S_e^* S_e = P_{t(e)}  \ \ \ \text{ and } \ \ \  P_v = \sum_{e: s(e)=v} S_e S_e^*,
\end{equation}
for all edges $e\in E^1$ and for all vertices $v\in E^0$ that are not sinks.

\smallskip

Given $\lambda_e \in (0,1)$, we consider as in \cite{CMR}
a quantum statistical mechanical system on the graph $C^*$-algebra $C^*(\cT/\Gamma)$, 
with time evolution determined by
\begin{equation}\label{evolCKgraph}
\sigma_t(S_e) = \left\{ \begin{array}{ll}
\lambda_e^{it} S_e & e\in G_\Gamma^1\\[2mm]
S_e & e\notin G_\Gamma^1.
\end{array}\right.
\end{equation}
The time evolution acts trivially on the trees and nontrivially (by a gauge action)
on the finite graph $G_\Gamma=\cT_\Gamma/\Gamma$.

\smallskip

In this setting, the notion of KMS state is generalized to the notion of KMS 
{\em weight}: these are positive norm lower semi-continuous functional $\phi$
on the graph algebra $C^*(E)$ that are invariant under the gauge action
and satisfy the KMS condition $\phi(ab)=\phi(\sigma(b)a)$, where
$$ \sigma(S_\mu S_\nu^*) = \frac{\lambda(\nu)}{\lambda(\mu)} S_\mu S_\nu^*, $$
for multi-indices $\mu$ and $\nu$, with $\lambda(\mu)=\prod_j \lambda(e_j)$
for $\mu=e_1\cdots e_k$. It is shown in \cite{CMR} that (faithful) KMS weights are in
one-to-one correspondence with (faithful) {\em graph weights}.

\smallskip

A {\em graph weight} is a pair of functions $g: E^0 \to \R_+$ and $\lambda: E^1 \to \R_+$
satisfying the conservation equation at vertices $v\in E^0$
\begin{equation}\label{graphweight}
g(v) = \sum_{s(e)=v} \lambda(e)\, g(t(e)).
\end{equation}
It is {\em faithful} if $g(v)\neq 0$ for all $v\in E^0$.

\smallskip

As in \cite{CMR}, we focus in particular on the case of {\em special graph weights}
where $\lambda(e)=\lambda^{n_e}$ for a fixed $\lambda\in (0,1)$ with $n_e=0$
for $e\notin G_\Gamma^1$ and $n_e=1$ for $e\in G_\Gamma^1$. These correspond
to KMS weights for the time evolution as in \eqref{evolCKgraph} with 
$\lambda_e=\lambda^{n_e}$, see \S 4 of \cite{CMR}.

\smallskip

Moreover, it is shown in \cite{CMR} that one can always construct a special
graph weight for the graph $E=\cT/\Gamma$ in the following way. Let
$n=\#G_\Gamma^0$ and order these vertices so that the first $r$ of them
(with $r\leq n$) are not sinks in $G_\Gamma$, while the remaining $n-r$
are sinks (though they are not sinks in the larger graph $\cT/\Gamma$.
Then consider the matrix
$A_{G_\Gamma}=(A_{G_\Gamma})_{ij}$ with $i,j=1,\ldots, n$,
given by
\begin{equation}\label{AGmatrix}
A_G = \left( \begin{array}{cc} \lambda m_{ij} & \lambda m_{ik} \\ 0 & 1 
\end{array}\right)
\end{equation}
where in the left top $r\times r$-block has entries $m_{ij}$ given by
the number of edges of $G_\Gamma^1$ connecting the vertices $v_j$ and $v_j$
and the right top $r \times (n-r)$-block has entries $m_{ik}$ equal to the
number of edges from $v_i$ to a sink $v_k$, while the bottom left $(n-r)\times r$-block
consists of zeros and the right bottom $(n-r)\times (n-r)$-block is the identity matrix.
This has eigenvalue is $\lambda_G=1$ with an eigenvector 
$v_G$ with positive entries $v_{G,i} >0$. Then setting 
\begin{equation}\label{vGi}
g(v_i)=v_{G,i} 
\end{equation}
gives a solution of the graph weight equations
\begin{equation}\label{specialgw}
 g(v_i)=\sum_{j=1}^n \lambda m_{ij} g(v_j), \ \  i\leq r \ \ \text{ and } \ \ 
g(v_i) = \sum_{j=1}^n \delta_{ij} g(v_j), \ \  r<j\leq n. 
\end{equation}
A graph weight obtained in this way on the finite graph $G_\Gamma=\cT_\Gamma/\Gamma$ extends
to a graph trace on the attached trees in $E=\cT/\Gamma$, by propagating it along
the trees using a solution of the equation
\begin{equation}\label{traceeq}
 g(v) = \sum_{s(e)=v} g(t(e)). 
\end{equation}

\smallskip
\subsection{A stochastic process for eternal inflation}

We now show that we can use the construction of faithful graph weights
recalled above from \cite{CMR} provides us with a model for a stochastic
process on the graph $E=\cT/\Gamma$ that replaces the stochastic process
on the Symmetree.

\smallskip

We have seen that, in the case of an eternal inflation model on a tree, we can construct a stochastic
process by setting 
\begin{equation}\label{PhimuWtrees}
\Phi_a(m+1) = \sum_{w\in \cW_{\fA,m}} \mu_{W,\beta}(\Lambda(wa)) =\sum_{w\in \cW_{\fA,m}}
\varphi_\beta(S_{wa}S^*_{wa}),
\end{equation}
where $\mu$ is a measure on $\Lambda_{\fA}$ that corresponds to the KMS
state $\varphi_\beta$ on $\cO_{\fA}$ for the time evolution associated to the Keane
potential $W$. In the case of an eternal inflation model on a quotient $\cT/\Gamma$,
we need to show that it is possible to ``interpolate" these constructions on the trees
sticking out of the vertices of $G_\Gamma=\cT_\Gamma/\Gamma$ in a consistent way across the
finite graph $G_\Gamma$.

\smallskip

Notice that, if we want a good notion of a stochastic process generalizing \eqref{PhimuWtrees}
to a graph that is not a tree, we need to take into account the fact that the probability
$\Phi_a(m)$ of reaching a certain state $a$ after $m$ steps will depend on the choice
of the intermediate vertices visited along the path (unlike in a tree where the choice is unique).
This is clear, since different chains of vertices will present different branching possibilities
for the process, hence will affect the resulting probability. Thus, we expect to have
a process of the form $\Phi_{\underline{v}, a}(m)$, where $\underline{v}=(v_1, \ldots, v_{m-1})$
is the sequence of vertices of the graphs visited by a path of length $m$. There will be 
several paths with the same sequence of vertices if the graph has multiple edges. (We
are allowing this possibility, see the definition of the matrix $A_G$ above.)

\smallskip

In the following, for an oriented path $\gamma=e_1 \cdots e_m$ of edges 
in $\cT/\Gamma$, we let $|\gamma|_\sigma$ be defined as in \cite{CMR} 
as $|\gamma|_\sigma = \sum_j n_{e_j}$ so that 
$$ \lambda(\gamma) = \prod_j \lambda(e_i) = \lambda^{\sum_j n_{e_j}} =
\lambda^{|\gamma|_\sigma}. $$

\smallskip

The following result shows that graph weights provide a way to construct a
stochastic process on $\cT/\Gamma$ with the desired properties.

\smallskip

\begin{thm}\label{weightprocess}
Let $\cT_i$ be the trees satisfying 
$\sqcup_i (\cT_i \smallsetminus \{ v_i \})= (\cT/\Gamma) \smallsetminus (\cT_\Gamma/\Gamma)$,
with $v_i$ the root vertex of $\cT_i$. Let $\Lambda_i =\partial \cT_i$ be the boundaries of these
trees, $\Lambda_i \subset X_\Gamma(\bK)$. Suppose given measures $\mu_i$ on each 
$\Lambda_i$ satisfying $\mu_i(\Lambda_i)=v_{G,i}$, with $v_G$ as in \eqref{vGi}, and
absolutely continuous with respect to the uniform Hausdorff measure of dimension
$\dim_H(\Lambda_i)$.  For $v$ a vertex of $\cT_i$ let $\Lambda_i(v)\subset \Lambda_i$ 
be the boundary of the subtree of $\cT_i$ with root $v$.
Then there exists a special graph weight $(g,\lambda)$ on $\cT/\Gamma$
satisfying \eqref{specialgw} on the finite
graph $\cT_\Gamma/\Gamma$ and equal to $g(v)= \mu_i(\Lambda_i(v))$ on $\cT_i$.
Let $\phi_{g,\lambda}$ be the associated KMS weight on the graph algebra
$C^*(\cT/\Gamma)$. Then consider 
\begin{equation}\label{Phiagraph}
 \Phi_{\underline{v},a}(m+1)=\sum_\gamma \phi_{g,\lambda}(S_{\gamma a}
S^*_{\gamma a}) =\sum_\gamma \lambda^{|\gamma a|_\sigma} g(r(\gamma a)), 
\end{equation}
with the sum over all oriented paths $\gamma$, of length $\ell(\gamma)=m$, 
passing through the given sequence
of vertices $\underline{v}$, and completely contained inside $\Delta_\Gamma/\Gamma$.
The probabilities $\Phi_{\underline{v},a}(m+1)$ can be equivalently written as
\begin{equation}\label{Phiagraph2}
 \Phi_{\underline{v},a}(m+1)= A_{G,v_0 v_1} A_{G, v_1 v_2} \cdots A_{G v_m v_{m+1}} 
g(v_{m+1}) 
\end{equation}
where $A_G$ is the matrix \eqref{AGmatrix} and $g(v_{m+1})=g(r(\gamma a))$.
This determines a stochastic process on $\cT/\Gamma$ that consistently
extends to the finite graph $\Delta_\Gamma/\Gamma$, with stochastic processes
\eqref{PhimuWtrees} on the trees, associated to the measures $\mu_i$.
\end{thm}

\proof We construct a special graph weight $(g,\lambda)$  on $\cT/\Gamma$
as in \cite{CMR}, by a solution of the equation \eqref{specialgw} on the finite
graph $\cT_\Gamma/\Gamma$. If $v_i$ is a vertex of $\cT_\Gamma/\Gamma$
that is a root of a tree $\cT_i$ in $\cT/\Gamma$, the value of $g$ at $v_i$ is
given by $g(v_i)= v_{G,i}$, where $v_G$ is the normalized (right) Perron--Frobenius eigenvector of
the matrix $A_G$ of \eqref{AGmatrix}. In order to extend $g$ to the tree $\cT_i$
we need to propagate the value $g(v_i)=v_{G,i}$ at the root vertex to the rest
of the tree, using a solution of \eqref{traceeq}. We obtain such a solution by
setting $g(v)= \mu_i(\Lambda_i(v))$, for $v$ a vertex of $\cT_i$. 
By the additivity of the measure $\mu_i$ this satisfies
$$ \mu_i(\Lambda_i(v)) = \sum_{s(e)=v} \mu_i(\Lambda_i(t(e))). $$
The absolute continuity condition on the measures $\mu_i$ implies that
$\mu_i(\Lambda_i(v))\neq 0$ for all $v$, hence we obtain a special graph
weight $(g,\lambda)$ as in \cite{CMR}. 

By Lemma 5.2 of \cite{CMR}, the sum $\phi_{g,\lambda}(S_{\gamma a}S^*_{\gamma a})$
can be written as 
$$ \sum_\gamma \lambda^{|\gamma a|_\sigma} g(r(\gamma a)) =  M_{\underline{v}} \, 
\lambda^{m+1} g(r(\gamma a)), $$
where $M_{\underline{v}}$ is the number of paths going through the specified
sequence of vertices
$$ M_{\underline{v}} = \prod_i m_{v_i v_{i+1}}, $$
with the multiplicities $m_{ij}$ as in the matrix $A_G$ of \eqref{AGmatrix}. Since all
the edges are in the finite graph $\cT_\Gamma/\Gamma$, we have $n_{e_j}=1$ for
all edges in the path, hence $\lambda^{|\gamma a|_\sigma}=\lambda^{m+1}$. Therefore,
we can identify
$$ M_{\underline{v}} \, \lambda^{m+1} g(r(\gamma a)) =A_{G,v_0 v_1} A_{G, v_1 v_2} \cdots A_{G v_m v_{m+1}}  g(r(\gamma a)). $$ 
\endproof

\begin{rem}\label{spflowrem} {\rm
The values $\phi_{g,\lambda}(S_{\gamma}
S^*_{\gamma}) = \lambda^{|\gamma|_\sigma} g(r(\gamma))$
have an interpretation as a spectral flow, as explained in \cite{CMR},
with respect to the unbounded operator given by the
infinitesimal generator of the time evolution 
(see Lemma 5.2 of \cite{CMR}).}\end{rem}

\smallskip

{}From the point of view of the eternal inflation model, this creates a scenario
where the evolution allows for trajectories that remain confined within the
bounded region $G_\Gamma=\cT_\Gamma/\Gamma$ while others escape this region
and, once they enter one of the infinite trees in $\cT/\Gamma$ outside of
$G_\Gamma$, they reproduce the behavior of the original
Eternal Symmetree model.

\smallskip
\subsection{Towards a continuum $p$-adic model?}

In $p$-adic geometry, there is a natural way to lift the discrete Bruhat--Tits tree
to a continuum model. The latter is provided by the Drinfeld $p$-adic upper half plane
$\H_\bK = \P^1_\bK \smallsetminus \P^1(\bK)$. This is a rigid analytic space with a projection
map to the Bruhat--Tits tree $\Lambda: \H_\bK \to \cT_\bK$. This map has the property that,
for any two vertices $v, w$ of $\cT_\bK$ connected by an edge $e$, the preimages 
$\Lambda^{-1}(v)$ and $\Lambda^{-1}(w)$ are open subsets of $\Lambda^{-1}(e)$,
see \cite{BouCar} for more details.

\begin{center}
\begin{figure}
\includegraphics[scale=.45]{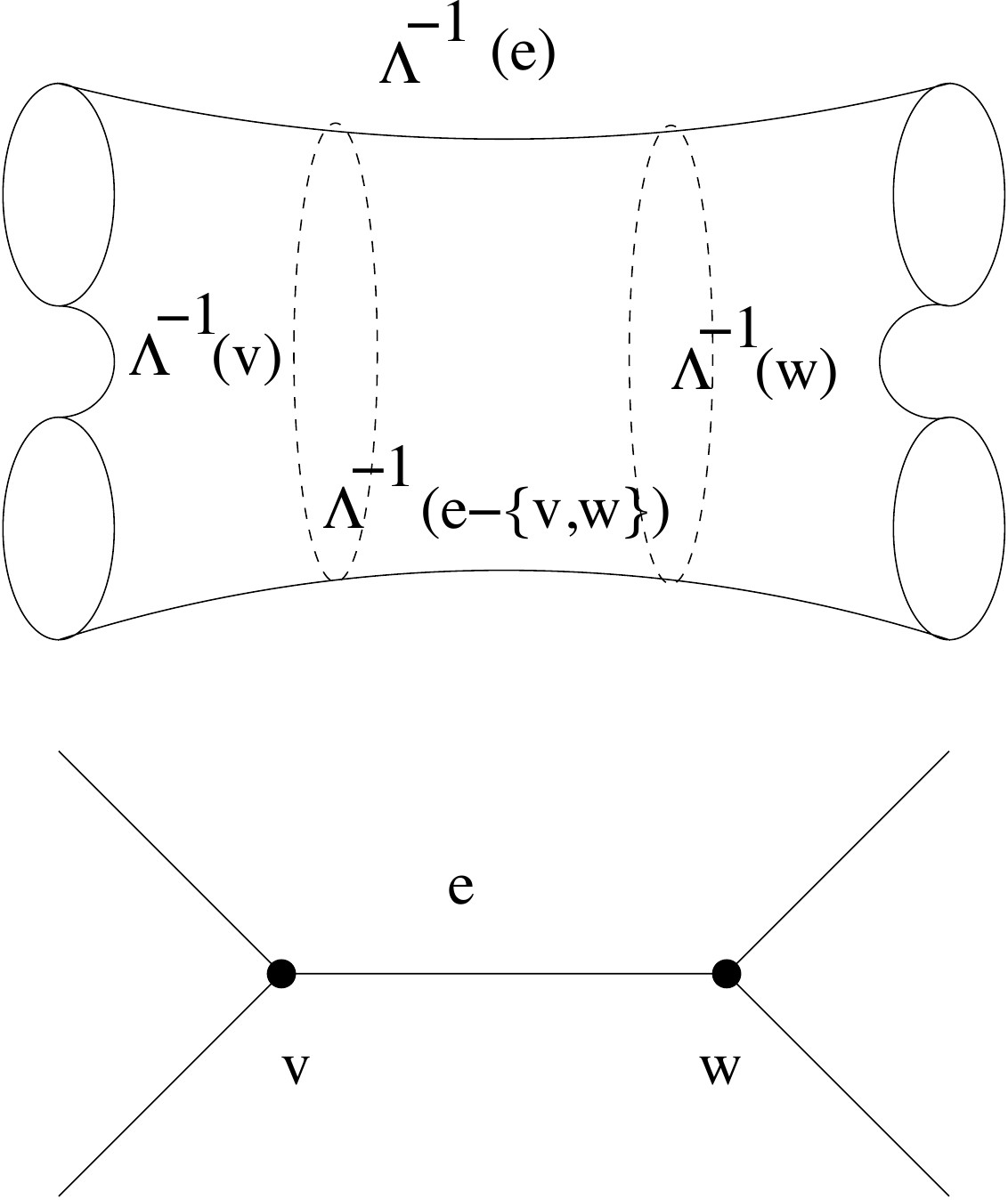}
\caption{Drinfeld's $p$-adic upper half plane and the Bruhat--Tits tree (from \cite{CMR}).}
\end{figure}
\end{center}

A possible way to lift the eternal inflation model of the Eternal Symmetree of \cite{HSSS} 
from the discrete level of the Bruhat--Tits tree to a continuous model living on the
$p$-adic upper half plane is to allow for more general stochastic processes given by
``signed measures", which in turn can be related, as shown in \cite{CMR}
to solutions of the graph weight equations and to theta functions on Mumford curves. 

\smallskip 

More precisely, let $\fm\subset \cO_\bK$ be the maximal ideal, with $\cO_\bK/\fm$
the residue field with $q=\#\cO_\bK/\fm$. Let $\pi$ be a uniformizer, namely $\fm=(\pi)$.
Let $|\cdot |$ be the absolute value with $|\pi |=q^{-1}$. 
The spectral norm of a function on the $p$-adic upper half plane
is defined as 
$$ \| f \|_{\Lambda^{-1}(v)} = \sup_{z\in \Lambda^{-1}(v)} |f(z)| . $$

\smallskip

The following fact is shown in \cite{CMR}. We recall it here for the reader's convenience.

\begin{lem}\label{thetaweight}
Given theta function $f\in \Theta(\Gamma)$, setting 
\begin{equation}\label{gvtheta}
 g(v)= \log_q \| f \|_{\Lambda^{-1}(v)}
\end{equation}
determines a function $g: \cT_\bK^0 \to \Z$ satisfying the graph weight equation
\begin{equation}\label{weighteqtheta}
 g(v) = \frac{1}{N_v} \sum_{s(e)=v} g(r(e)),
\end{equation} 
where $N_v=\# \{ e\,:\, s(e)=v \}$.
\end{lem}

\proof As shown in \cite{Put}, 
a theta function $f\in \Theta(\Gamma)$ on the Mumford curve $X_\Gamma$
determines a current $\mu$ on the graph $\cT_\bK/\Gamma$ given by the
growth of the spectral norm of $f$ in the $p$-adic upper half plane
$$ \mu(e) = \log_q  \| f \|_{\Lambda^{-1}(r(e))}  - \log_q  \| f \|_{\Lambda^{-1}(s(e))}, $$
with $r(e)$ and $s(e)$ the range and source of the oriented edge $e$.
The current satisfies a momentum conservation equation of the form
$$ \sum_{s(e)=v} \mu(e) =0. $$
This conservation equation for $\mu(e)=g(r(e))-g(s(e))$ in turn implies the 
stated graph weight equation for the function $g(v)$.
\endproof

This suggests that allowing for stochastic processes associated to
signed measures of total mass zero, instead of positive probability 
measures, may lead to an interesting connection between discrete
eternal inflation models on the $p$-adic Bruhat--Tits tree and
continuous lifts to the $p$-adic upper half plane related to
$p$-adic automorphic functions of the type considered in \cite{Man},
\cite{Put}.

\bigskip

\subsection*{Acknowledgment} The second author is supported by
a Summer Undergraduate Research Fellowship at Caltech and by 
the Rose Hills Foundation. The first author is supported by NSF grants
DMS-0901221, DMS-1007207, DMS-1201512, PHY-1205440.

\end{document}